\documentclass[pre,twocolumn,showpacs,preprintnumbers,amsmath,amssymb]{revtex4}

\usepackage{graphicx}
\usepackage{dcolumn}
\usepackage{bm}

\begin{document}

\newcommand{\beq}{\begin{equation}}
\newcommand{\eeq}{  \end{equation}}
\newcommand{\bea}{\begin{eqnarray}}
\newcommand{\eea}{  \end{eqnarray}}
\newcommand{\bit}{\begin{itemize}}
\newcommand{\eit}{  \end{itemize}}

\title{Quantum to classical transition in a system with a mixed classical
dynamics}

\author{Fabricio Toscano}

\affiliation{Funda\c c\~ao
Centro de Ci\^encias e Educa\c c\~ao Superior \`a Dist\^ancia do
Estado do Rio de Janeiro, 20943-001 Rio de Janeiro, RJ, Brazil.}

\affiliation{Instituto de F\'{\i}sica, Universidade Federal do Rio
de Janeiro, Cx. P. 68528,\\
21941-972 Rio de Janeiro, RJ, Brazil.}

\author{Diego A. Wisniacki}
\affiliation{Departamento de F\'{\i}sica ''J. J. Giambiagi'', FCEN, UBA,
1428 Buenos Aires, Argentina.}

\date{\today}

\begin{abstract}

We study how decoherence rules the quantum-classical transition of the
Kicked Harmonic Oscillator (KHO).
When the amplitude of the kick is changed the system presents a classical dynamics that
range from regular to a strong chaotic behavior. 
We show that for regular and mixed classical dynamics, 
and in the presence of noise, the distance between 
the classical and the quantum phase space distributions is proportional 
to a single parameter $\chi\equiv K\hbar_{\rm eff}^2/4D^{3/2}$ which relates 
the effective Planck constant $\hbar_{\rm eff}$, the kick amplitude
$K$ and the diffusion constant $D$. This is valid when 
$\chi < 1$, a case that is always attainable in the semiclassical regime
independently of the value of the strength of noise given by $D$.
Our results extend a recent study performed in the chaotic regime.

\end{abstract}

\pacs{05.45.-a,03.65.Ta}

\maketitle

\section{Introduction}

The complete description of the emergence of the classical world within the
quantum theory is a fundamental problem that have attracted a lot of attention  
since the beginning of quantum mechanics. 
Although many advances have been made, there are still open questions.
An important step forward was done with the understanding that the formalism of
phase space distribution functions is the appropriate framework to study
the quantum to classical transition \cite{Ballentine1994}.
Among all quantum phase-space distributions, the Wigner function 
\cite{Wigner1932} has the advantage that identifies the nonlinearities 
of the system as the responsible for the 
separation between the quantum and the classical evolution.
Indeed, while for linear systems the Wigner distribution obey the Liouville 
equation of the classical distribution, the nonlinearities add terms to this
equation that eventually set the two distributions apart \cite{Hillery1984}, 
even for initial states that are classically allowed.

The time for which classical and quantum distributions starts to 
differ, called breaking or separation time, can be very short for 
classically chaotic systems.
This is because the nonlinearities of the system are reached
very quickly due to the exponential stretching of the distributions.
Hence, the breaking time scale as $\ln(1/\hbar_{\rm eff})$, where 
the effective Planck constant $\hbar_{\rm eff}\equiv\hbar/S$ is the 
semiclassical parameter (with $S$ a typical
action of the system)
\cite{Berman1978,Berry1979,Chirikov1988,
Zaslasky1991,Karkuszewski2002,Andre2004,Toscano-proceedings}.

An important and subtle problem is that the breaking time
for macroscopic systems ($\hbar_{\rm eff}\ll 1$) can 
still be small compared to a typical evolution 
time due to its logarithmic behavior,
allowing quantum effects that are not observed in 
the classical world \cite{hyperion}. 
This paradoxical situation is explained by the coupling 
of the system with the environment, which leads to the elimination of the quantum signatures, providing the reconciliation of theory and observation 
\cite{Paz1994,Zurek2003}. At this respect, an open question 
is how the effects of the environment affect the logarithmic law of the breaking 
time in systems with a classical chaotic dynamics. 
More precisely, we can say that it is still unknown the 
exact relation between the strength of noise and the effective 
Planck constant in order to obtain the correct classical limit 
\cite{Saraceno2005}.


This problem has started to be answer in a recent work \cite{Toscano2005}
where the authors study the effect of a purely diffusive environment
over an specific model given by the Kicked Harmonic Oscillator (KHO).
They show that the differences between the classical and quantum phase evolution 
is proportional to a single parameter
$\chi=K\hbar_{\rm eff}^2/4D^{3/2}$, that
combine $\hbar_{\rm eff}$, the diffusion coefficient $D$ and the kicking strength $K$,
who control the macroscopicity, the noise and the chaotic behavior
respectively.
These quantum-classical differences were estimated with ${\cal D}_n$ ($n$ is the number of
kicks) which is the integral over the whole phase space of the modulus of the difference
between the quantum and the classical distributions.
Thus, it was shown that, when the classical dynamics is chaotic, in the semiclassical limit $\hbar_{\rm eff}\ll 1$ and when $\chi < 1$, the single parameter $\chi$ 
controls the quantum-classical transition of the KHO. 
This result, confirm the conjecture presented in Ref. \cite{Pat} that in the presence of noise 
a single parameter controls the quantum-classical transition of classically chaotic systems.

In this paper we analyze the case when the classical dynamics of the KHO is 
regular or mixed. 
In the absence of a reservoir, we show that the separation between
the quantum and the classical evolution can be characterized through 
the behavior of ${\cal D}_n$
as a function of time.
When the initial distribution explore regions of regular classical
dynamics, the behavior of ${\cal D}_n$ allows the 
identification of a potential scale-law
$1/\hbar_{\rm eff}^{{\alpha}}$ for the separation time.
However, this scale-law is not valid when the degree
of mixing increase.
In the presence of a purely diffusive reservoir, we show that the 
quantum-classical differences measured by ${\cal D}_n$, still scale 
with the parameter $\chi$ in the semiclassical limit and when $\chi < 1$.
Hence, the concept of separation time is not meaningful anymore, since the 
quantum an the classical distributions remain close throughout the evolution.
Our result suggest that the conjecture 
in \cite{Pat} can be extended independently of the underlying classical dynamics.
 
The paper is organized as follows. Section \ref{section1} is devoted to
review the quantum and the classical versions of the KHO and also to explain the 
formalism for the evolution of the Wigner function and the classical
distribution as a map in phase space. 
In Section \ref{section2} we introduce our measure of the quantum-classical
differences ${\cal D}_n$, and study its behavior as a function of time 
without the presence of a reservoir.
We investigate the scale-law with the effective Planck constant
of the separation time when the initial distribution explore regular or mixed 
regions of the phase space.
The effects of decoherence over ${\cal D}_n$, when the KHO is in contact with a purely
diffusive reservoir, is studied in Section \ref{section3}. 
We end with some final remarks in Section \ref{section4}.

\section{Model: The Kicked Harmonic Oscillator}
\label{section1}

\subsection{The quantum version}

The model used in our calculations is the KHO,
a particle of mass $m$ in a harmonic potential subjected to a periodically applied
position-dependent delta pulses.
The quantum Hamiltonian is defined as:
\begin{equation}
\label{Hamiltonian_kicked_rotor}
\hat{H}=\frac{\hat{P}^2}{2m}+\frac{1}{2}m\nu^2\hat{Q}^2+
A\;\cos(k\hat{Q})\sum_{n=0}^\infty \delta(t-n\tau),
\end{equation}
where $\nu$ is the oscillator frequency, $A$ the amplitude of the kicks
and $\tau$ the interval between two consecutive kicks.
$k$ sets the periodicity of the position dependent kicking potential.
The physical realization of this Hamiltonian is within the context of
trapped ions. It was shown in Ref. \cite{Gardiner1997}, that it describes
the center-of-mass dynamics
of an ion in a one-dimensional trap submitted to a sequence of
standing-wave laser-pulses, off-resonance with a transition between
the electronic ground state and another internal state.
The wave number $k$ in Eq.~(\ref{Hamiltonian_kicked_rotor}) is the projection
along the trap axis of the corresponding wave vectors with equal modulus
$|\vec{k}|$ of two opposite propagating pulses with oblique incidence
measured by an angle $\theta$, {\it i.e.} $k\equiv2|\vec{k}|\cos(\theta)$.

It is possible to define an effective Planck constant $\hbar_{\rm eff}$
if we work with dimensionless quantities
$\hat{q}=k\hat{Q}$, $\hat{p}=k\hat{P}/m\nu$, and $K=k^2A/m\nu$,
so that $[\hat{q},\hat{p}]=2i\eta^2\equiv i\hbar_{\rm eff}$, with
$\eta=k\Delta Q_0=k\sqrt{\hbar/2m\nu}$, and $\Delta Q_0$ being the width of the
ground state of the harmonic oscillator.
The dimensionless parameter $\eta$ is the so-called Lamb-Dicke
parameter \cite{Wineland}, which measures the ratio between the
ground state width and the wavelength $\lambda=2\pi/k$ that sets
the scale of the non-linearity of the Hamiltonian.
In experiments with trapped ions, the classical limit
$\eta \rightarrow 0$ (and thus $\hbar_{\rm eff}\equiv2\eta^2 \rightarrow 0$) can be approximated
simply by changing the angle $\theta$ of incidence of the incoming pulses,
or by increasing the trap frequency $\nu$.

\begin{figure}[t]
\setlength{\unitlength}{1cm}
\begin{center}
\includegraphics[width=8.5cm]{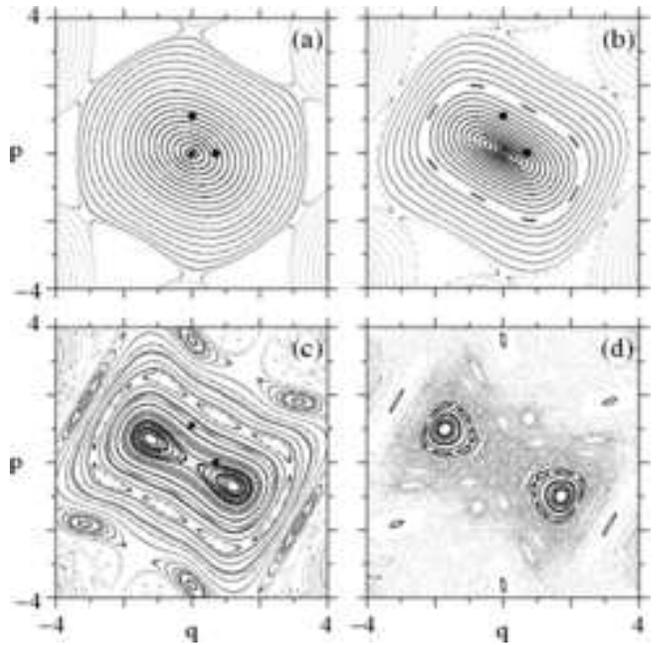}
\end{center}
\caption{Stroboscopic phase space of the KHO for
different values of the kicked amplitude $K$. In  {\bf (a)} $K=0.5$, {\bf (b)} $K=1.0$,
{\bf (c)} $K=1.5$ and in {\bf (d)} $K=2.0$. The phase space points $(q_0,p_0)=(0,0)$, $(0,1.1)$ and $(0.7,0)$, marked with symbols ($\bullet$), are the centers of
the initial coherent states used in our numerical simulations.}
\label{fig1}
\end{figure}

\subsection{The classical dynamics.}

The classical Hamiltonian is obtained from Eq.(\ref{Hamiltonian_kicked_rotor})
simply by replacing the quantum operators $\hat{Q}$ and $\hat{P}$ by the phase
space variables $Q$ and $P$ respectively. In an analogous way we can define
the dimensionless phase space coordinates ${\bf x}\equiv(q,p)\equiv(kQ,kP/m\nu)$.
Then, the time-dependent classical evolution can be described
as a composition of a discrete map corresponding to the kick, plus a rotation in
phase space:
${\bf x}_{n+1}={\bf R}\circ {\bf K}({\bf x}_{n})$ where
${\bf x}_n\equiv(q_n,p_n)$ are the coordinates before
kick $n$ (where the first kick corresponds to $n=0$).
The kick operation ${\bf K}$ is defined as
\begin{equation}
\label{K_operation}
q_{n}^+=  q_{n}\,,\qquad
p_{n}^+= p_{n} + K\,\sin(q_n)\,,
\end{equation}
and the phase space rotation ${\bf R}$ is given by:
\begin{eqnarray}
\label{R_operation}
 q_{n+1}&=& \cos(\nu\tau)\,q_{n}^+ + \sin(\nu\tau)\,p_n^+ \,,\nonumber\\
p_{n+1}&=& -\sin(\nu\tau)\,q_{n}^+ + \cos(\nu\tau)\,p_n^+\;.
\end{eqnarray}
In close relation with Ref. \cite{Toscano2005}, we also consider the
case where the interval between kicks $\tau$ and the period of the harmonic 
oscillator $T=2\pi/\nu$ are related by $\tau=T/6$.

The phase space of the system is unbounded and corresponds to a mixed
dynamics exhibiting stable islands surrounded by a stochastic web
along which the particle diffuses. The thickness of the web is controlled
by the chaoticity parameter $K=k^2A/m\nu$ which is the dimensionless kicked amplitude.
Hence, these chaotic regions broadens (shrinks) as $K$ increases (decreases).
In the case of $\tau=T/6$ considered, the stochastic web displays crystal hexagonal
symmetry \cite{Zaslasky1991,Andre2004}.
In Fig.\ref{fig1} we show the stroboscopic phase space around the origin for
different values of $K$.
For $K\le K_c\approx 1.1547$ the origin is an elliptic fixed point while 
for $K > K_c$ the system undergoes a bifurcation where the origin become
an hyperbolic fixed point and two neighbor elliptic fixed points arise.
The separation between these two elliptic fixed points increases with $K$.
As we can see in Fig.\ref{fig1}(a), for $K=0.5$ the dynamics of the system around
the origin is essentially regular with a series of concentric tori around the 
elliptic fixed point.
For $K=1$ [Fig \ref{fig1}(b)], some tori were destroyed, and tiny chaotic regions appear (invisible to the eye).
Mixed dynamics star to manifest most appreciable for $K=1.5$ while
for $K=2$ most of the area around the origin shows chaotic behavior 
[see Fig .\ref{fig1}(c) and (d)].

\subsection{The evolution of phase space distributions without reservoir.}

We analyze the quantum-classical transition in phase space comparing the
evolution of the Wigner function \cite{Wigner1932} and the 
corresponding classical distribution.
The Wigner function, defined as
\begin{equation}
\label{def-wigner-function}
W({\bf x})\equiv \int dy \left\langle q+y/2
\left|\frac{\hat{\rho}}{4\pi\eta^2}\right|q-y/2\right\rangle \;
\exp\left\{-i\frac{py}{2\eta^2}\right\}\,,
\end{equation}
with $\hat{\rho}$ the density operator of the quantum state,
have a highlight position within all the quantum quasi-probability
distribution in phase space because it is the only one to yield 
the correct marginal probabilities, thus satisfying an important 
property of classical distributions \cite{Bertrand}.
Furthermore, quantum effects are more pronounced, as compared to other bounded
distributions. Finally, new methods have been
developed for the direct measurement of the Wigner function, which are
adequate for the description of evolving systems \cite{direct-measurement}.

Without the action of a reservoir the unitary evolution of the Wigner function
of the KHO can be written as \cite{Toscano2005,Berry1979}
\begin{equation}
\label{Wigner_evolution}
W_{n+1}({\bf x})=
\int d{\bf x'}\;L({\bf x}^R,{\bf x'})\;W_n({\bf x'})
\,\,,
\end{equation}
where $W_{n+1}$ and $W_n$ are the Wigner functions immediately before
the kicks $n+1$ and $n$ respectively. The integration is over all the unbounded
phase space, and
\begin{eqnarray}
\label{quantum-propagator}
L({\bf x}^R,{\bf x'})&\equiv&
\int_{-\infty}^{\infty} \frac{d\mu}{2\pi \eta^2}\;
e^{\frac{i}{\eta^2}\left[K \sin(q')\sin(\mu)-
\mu(p^R-p')\right]}\,\times \nonumber \\
&&\times\delta\left(q^R-q'\right)\,,
\end{eqnarray}
is the quantum propagator of one kick plus the harmonic evolution
between the consecutive kicks which is given by
\begin{equation}
\label{x-rotado}
{\bf x}^R\equiv[q^R({\bf x}),p^R({\bf x})]\equiv
{\bf R}^{-1}({\bf x})\;,
\end{equation}
{\it i.e.} the phase space coordinates rotated with the inverse transformation
of Eq.~(\ref{R_operation}).
The Liouville evolution of the classical distribution
$W^{cl}_n({\bf x})$ can also be written in the form of
Eq.~(\ref{Wigner_evolution}) with the classical propagator
\begin{equation}
\label{classical-propagator}
L^{cl}\left({\bf x}^R,{\bf x'}\right)\equiv
\delta\left[p'-p^R+K\sin(q')\right]\,\delta\left(q^R-q'\right)\;.
\end{equation}
In the classical limit $\eta\rightarrow 0$ the classical propagator is formally recovered from
the quantum one \cite{Berry1979}.
Indeed, in the semiclassical regime ($\eta\ll 1$) the integral in
Eq.(\ref{quantum-propagator}) can be estimated employing stationary-phase techniques
yielding,
\begin{eqnarray}
L({\bf x}^R,{\bf x'})&\approx&
\label{semiclassical-propagator}
\frac{1}{|b|^{1/3}}
\mbox{Ai}
\left(-\frac{\mbox{sign}(b)}{|b|^{1/3}} (p'-p^R+K\sin(q')) \right)
\,\times \nonumber \\
&&\times\delta\left(q^R-q'\right)\,,
\end{eqnarray}
where $b=(\eta^4 K/2)\sin(q')$ and $\mbox{Ai}(x)$ is the Airy function.
When $\eta \rightarrow 0$ we can use that
$\mbox{Ai}\{-y/\varepsilon\}/\varepsilon
\stackrel{\varepsilon\rightarrow 0}{\longrightarrow}
\delta(y)$ to get the classical Liouville propagator in
Eq.(\ref{classical-propagator}).
However, no matter how small we can do the Lamb-Dicke parameter
$\eta$ in a physical realization of the KHO, this will always take a finite value
and then the presence of an Airy function in the quantum propagator will eventually
set the Wigner function and the classical distribution apart.

\section{The quantum-classical separation without reservoir: scale-law
for the separation time.}
\label{section2}

The Wigner distribution of a linear system (quadratic Hamiltonian) satisfy
the Liouville equation for the classical phase-space probability distribution.
Thus, if we start with a Wigner function
of a classically allowed probability distribution (the only possibility is a
minimum uncertainty initial Gaussian wave packet \cite{Hudson1974}),
the linear dynamics evolution will preserve its Gaussian shape.
When the system have non-linearities,  the initial wave packet will deform and eventually delocalized during the evolution.
As a consequence, quantum interference effects appear
between different pieces of the wave packet,
which remains coherent throughout the unitary evolution.
These quantum interferences effects stem from the extra terms that the non-linearities of the system add to the Liouville equation in order to obtain the correct equation of motion of the Wigner function \cite{Moyal1949,Hillery1984}.
Equivalently, we can say that these interferences are originated
by the non-local nature of the Wigner propagator  
(see for example Eq.(\ref{quantum-propagator})] in comparison with the local nature of
the Liouville propagator of Eq.(\ref{classical-propagator}), that
persist even in the semiclassical regime [see Eq.(\ref{semiclassical-propagator})].
Hence, they are the responsible for the separation of the Wigner function from the classical distribution and thus of the quantum-classical correspondence breakdown.

The quantum-classical differences
start to become important when the initial distribution spreads over distances
of the order of the characteristic scale of the nonlinearities of the system,
and it is well known this argumentation in order to estimate the time scale for
the separation in the case of classically chaotic systems \cite{hyperion,Monteoliva2001,Zurek2003}.
Indeed, as the initial Gaussian Wigner function is well localized, it
will initially evolve following classical trajectories in phase space, therefore
the wave packet will stretch exponentially fast along the unstable manifold and
will shrinks also exponentially fast along the stable one in order to conserve
the phase space volume.
This is also accompanied by the characteristic folding of classical trajectories
that forces the wave packet to develops quantum interference
between different pieces which are coherently related.
As a result of this behavior, it is found a logarithmic scale law
with the effective Planck constant $\hbar_{\rm eff}$ 
of the separation time in systems with classical chaotic dynamics.
For the KHO, the estimation of the separation time $t_s$ gives
\begin{equation}
\label{separation-time-chaotic}
t_s=n_s\tau \approx \frac{\tau}{\Lambda}\ln(1/\eta)
\end{equation}
where $\Lambda$ is the local expansion coefficient and
$\hbar_{\rm eff}=2\eta^2$. A detailed estimation of $t_s$ can
be found in Ref. \cite{Zaslasky1991,Toscano-proceedings} and 
in \cite{Andre2004},
where it is also shown a numerical test that indicates that this is
also the time when quantum and classical expectation values start to 
differ from each other.
The expansion coefficient $\Lambda$ corresponds to the Lyapunov exponent
when an average over initial conditions in the chaotic regions is considered
\cite{Toscano2005,Toscano-proceedings}.

\begin{figure}[t]
\setlength{\unitlength}{1cm}
\includegraphics[width=7.5cm]{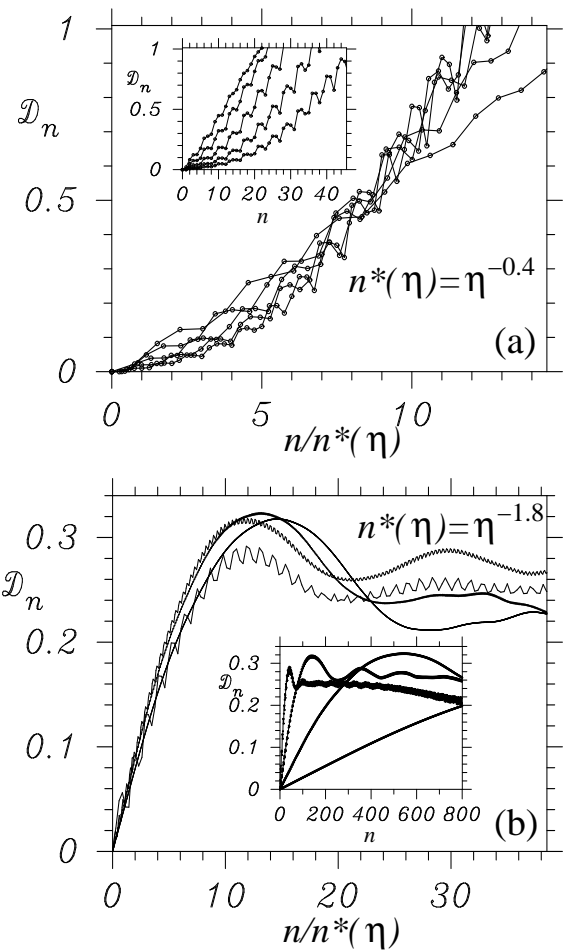}
\caption{Separation ${\mathcal D}_n$ between quantum and classical distribution
as a function of the number of kicks for $K=0.5$.
The full lines joint all the values of ${\mathcal D}_n$ calculated for the
same value of the Lamb-Dicke parameter $\eta$. In {\bf (a)} the initial
distribution is a coherent state centered at the phase space point
$(q0,p0)=(0,1.1)$ and the curves are for the decreasing values of
$\eta=0.5, 0.25, 0.125, 0.0625$ and $0.03125$. The result for the initial
condition $(q0,p0)=(0.7,0)$ is equivalent (not plotted here).
In {\bf b)} the initial coherent state is centered at the origin
and the curves are for $\eta=0.5, 0.25, 0.125$ and $0.0625$.
The number of kicks $n$ are rescaled by a function
$n^*(\eta)=\eta^{-\alpha}$ where we optimize the values of $\alpha$ in order
to obtain the best collapsing of all the curves.
The insets show the curves without the rescaling over the time $n$.}
\label{fig2}
\end{figure}

In the other extreme situation, when the classical dynamics of the
system is essentially regular, the scaling law of the quantum-classical separation
time with $\hbar_{\rm eff}$ was less studied.
The logarithmic dependence of $t_s$ in the chaotic case 
is due to the exponentially divergence of classical trajectories, but
in the integrable case the divergence is typically polynomial
\cite{Henon1981,Zurek2003}.
Therefore, following a similar type of argumentation that in the chaotic cas e,
the separation time in regular systems would follow a potential scale-law with
$\hbar_{\rm eff}$.
In the case of the KHO, if the initial distribution explores regions of essentially
regular dynamics, we would have
\begin{equation}
\label{separation-time integrable}
t_s=n_s\tau\approx \tau/\eta^{\alpha}\;,
\end{equation}
where numerical constants are forbidden.
Indications of the validity of Eq.(\ref{separation-time integrable}) for $K\ll 1$ are given in Ref. \cite{Zaslasky1991}.
In what follows we show a way to estimate the amount of quantum effects that lead to the separation between the Wigner function and the classical distribution,
and how the scaling law for the separation time in Eq.(\ref{separation-time integrable}) 
manifests in our formalism.

\begin{figure}[t]
\setlength{\unitlength}{1cm}
\includegraphics[width=7.5cm]{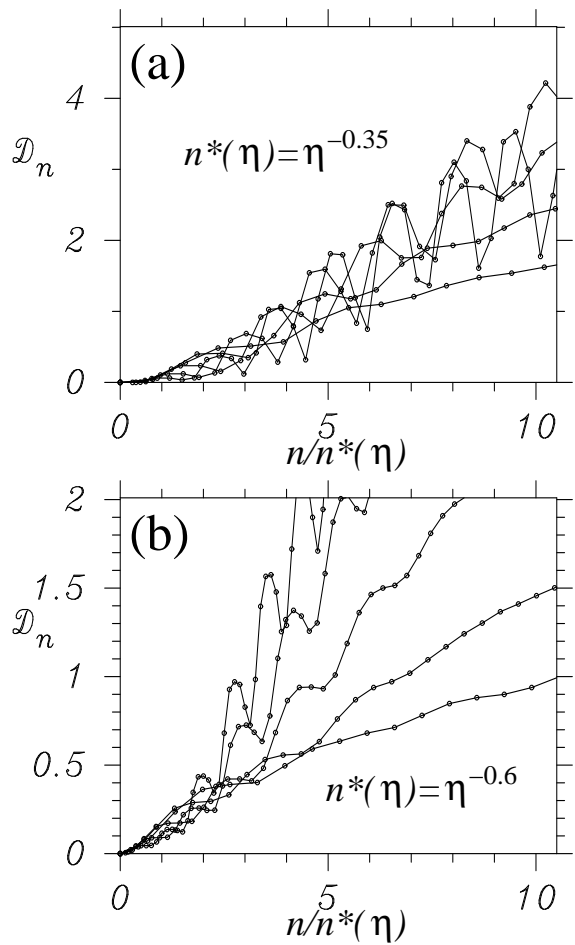}
\caption{Idem as Fig.{\ref{fig2}} but for $K=1.0$. In {\bf a)} the initial
coherent state centered at $(q0,p0)=(0,1.1)$  and in {\bf b)} $(q0,p0)=(0.7,0)$.
The different curves in each plot corresponds to
$\eta=0.5, 0.25, 0.125, 0.0625$ and $0.03125$.
The number of kicks $n$ are rescaled by a function
$n^*(\eta)=\eta^{-\alpha}$ where we optimize the values of $\alpha$ in order
to obtain the best collapsing of all the curves.}

\label{fig3}
\end{figure}

In Ref. \cite{Toscano2005}, it was introduced the time dependent quantity
($t=n\tau$),
\begin{equation}
\label{functional_distance}
{\mathcal D}_n\equiv
\int d{\bf x}
\left|W_{n}({\bf x})-W_{n}^{cl}({\bf x})\right|\;,
\end{equation}
to quantify the quantum effects that lead to the separation between
the Wigner function and the classical distribution.
Here, $W_n$  and $W_n^{cl}$ are
respectively the Wigner function and the classical distribution
immediately before the kick $n$, both normalized to the unity.
Starting with the same initial distribution it is expected
that during the unitary evolution the appearance of quantum-classical differences
make  ${\mathcal D}_n$ to grow and eventually saturate around some
finite value.
This behavior is shown in Fig.\ref{fig2} and {\ref{fig3}
where we have plotted ${\mathcal D}_n$ as a function of the scaled number of kicks
(see below) for
different values of the semiclassical parameter $\eta$ and the amplitude of
the kick $K$.
The initial distribution is always a coherent state with width
$\Delta q(0)=\Delta p(0)=\eta$.

In Fig.\ref{fig2} the amplitude of the kick is $K=0.5$, so we study the case with regular
classical dynamics as the different initial coherent states considered, during the
interval of times monitored, spread over the regular region around the origin
(see Fig.~\ref{fig1}, \ref{fig4} and \ref{fig5}).
We can see in Fig.~\ref{fig2} (a) that when we start with an
initial wave packet outside the origin, the quantity ${\mathcal D}_n$
grows up to values larger that one, for all values of
the semiclassical parameter $\eta$ \cite{comentario1}.
This clearly indicates the separation between the
Wigner and the classical distributions because a measure of their
differences reach values of the order of their normalization.
However, it is important to note that no matter which value of the quantum
differences, measured by ${\mathcal D}_n$, we choose in the plot of
Fig.~\ref{fig2} (a) to define the separation between the Wigner and
classical distributions: it will be reached at some early time of the
evolution independently of how small is the semiclassical parameter
$\eta$.
Thus, the separation time defined in this way should scale as in
Eq.(\ref{separation-time integrable}).
This is showed in the plots by the rescaling of the number of kicks
by a function of the form $n^*(\eta)=\eta^{-\alpha}$ that
allows the approximately collapse of all the curves.
In Fig.~\ref{fig3} it is shown the case when the kicking amplitude is
$K=1$.
We can see a similar behavior although the collapsing of all
the curves do not work so well.
This is an indication of the departure of the separation time
from the scale-law in Eq.(\ref{separation-time integrable}) when the degree of
mixing in the classical dynamics grows. It is important to note that such  
scaling was not observed for the case when the kicking amplitude is
$K=1.5$.

\begin{figure}[t]
\setlength{\unitlength}{1cm}
\begin{center}
\includegraphics[width=7.5cm]{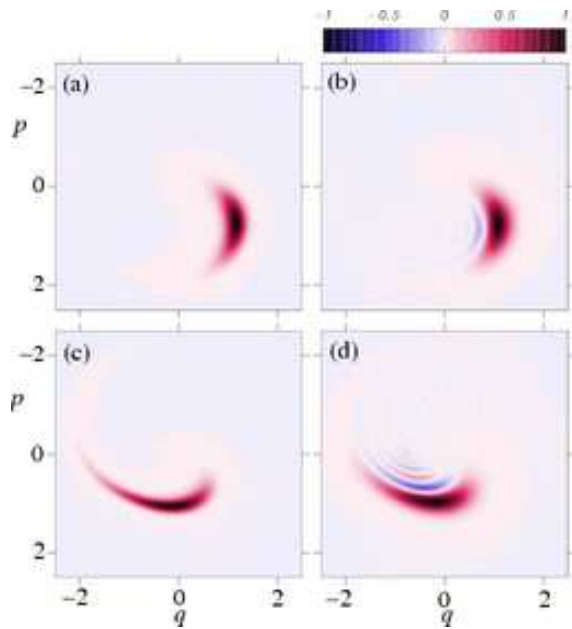}
\end{center}
\caption{Density plots of the Wigner (right)
and classical distributions (left)
that result from the evolution with the KHO Hamiltonian
with the kick amplitude $K=0.5$ from an initial coherent state
centered in the phase space point $(q_0,p_0)=(0.0,1.1)$
and with a width $\Delta q(0)=\Delta p(0)=\eta=0.25$.
In (a) and (b) immediately before the kick $n=8$ and
in (c) and (d) immediately before the kick $n=18$.}
\label{fig4}
\end{figure}

In Fig. \ref{fig2} (b) we show the case when the initial coherent state
is centered at the origin of phase space when $K=0.5$.
${\mathcal D}_n$ grows up to values
around $0.25$ and then saturate for all semiclassical parameters
$\eta$.
In this case, ${\mathcal D}_n$ grow less than when the
initial coherent state is outside the origin because  now the
quantum and classical distributions remains well localized for
very long times and thus the quantum-classical differences are less
pronounced (Fig.~\ref{fig5}).
Nevertheless, this differences do not decrease for small
values of $\eta$, showing that the separation
between the classical and the quantum evolution occurs in
the semiclassical regime.
The fluctuations observed in Fig.~\ref{fig2} (a) and (b)
have a period of $n\approx 6$ kicks {\it i.e.}
approximately the period of the harmonic evolution $T=6\tau$.
This is the time that most parts of the quantum and classical distribution
take to complete a spin along the elliptical structure of phase space around the origin for $K=0.5$ (see Fig.~\ref{fig1}).

It is interested to compare in Fig.~\ref{fig4} and Fig.~\ref{fig5} the
quantum-classical differences that leads to the separation between the
quantum and the classical evolution when the amplitude of the kick is $K=0.5$.
Indeed, when the initial coherent state is centered outside the origin,
the quantum and classical distributions spread out and as a consequence
the Wigner function develops a quantum interference pattern throughout the evolution
that clearly indicates its separation from the classical distribution
(see Fig.~\ref{fig4}).
Now, when the initial coherent state is centered at the origin of phase space,
which is an elliptical fixed point for $K=0.5$, the quantum and
the classical distributions remain well localized for very long times (Fig.~\ref{fig5}).
In this case, the classical distribution develops the typical classical structure
of ``whorls" associated with an elliptic fixed point \cite{Berry1979,Korsch1981}.
It is evident from the figure that the quantum evolution smooths out this
details of the classical distribution and this is the origin
of the quantum-classical differences
that leads to the separation.

\begin{figure}[t]
\setlength{\unitlength}{1cm}
\begin{center}
\includegraphics[width=7.5cm]{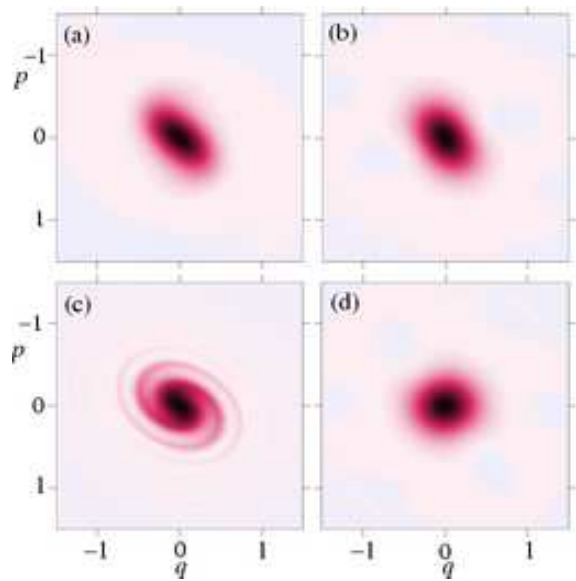}
\end{center}
\caption{Density plots of the Wigner (right)
and classical distributions (left) that result from the evolution with the KHO
Hamiltonian with the kick amplitude $K=0.5$ from an initial coherent state
centered at the origin of phase space and with a
width $\Delta q(0)=\Delta p(0)=\eta=0.25$.
In (a) and (b) immediately before the kick $n=60$ and
in (c) and (d) immediately before the kick $n=300$.}
\label{fig5}
\end{figure}



\section{Decoherence: the effect of a diffusive environment}
\label{section3}

%
\begin{figure*}[t]
\setlength{\unitlength}{1cm}
\includegraphics[width=14.5cm]{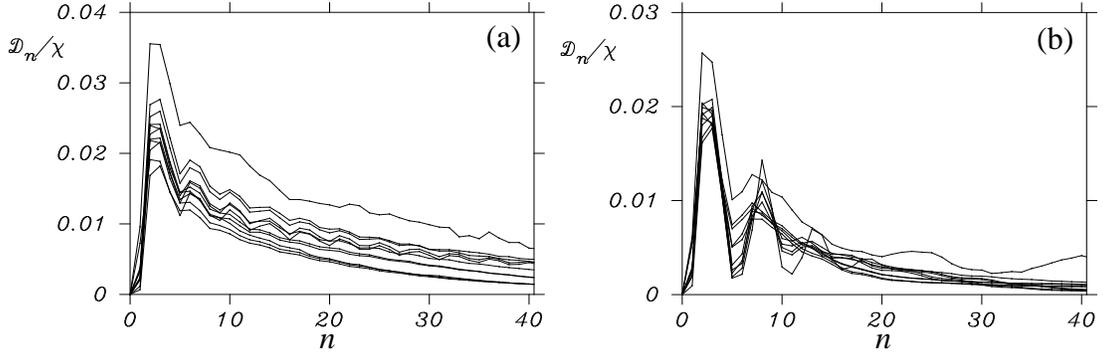}
\caption{Renormalized distances between quantum and classical distributions,
with diffusion. All the curves were generated with the initial coherent state
centered in the phase-space point $(q_0,p_0)=(0,1.1)$ (see Fig.~\ref{fig1} )
with  the kicked amplitude $K=0.5$ in {\bf (a)} and
$K=1.5$ in {\bf (b)}.
The eleven different curves correspond
to $\chi\equiv K\eta^4/D^{3/2}=6.2\times10^{-2},4.3\times10^{-2},1.5\times10^{-2},
1.1\times10^{-2},
7.6\times10^{-3},3.9\times10^{-3}, 1.4\times10^{-3},
6.8\times10^{-4}, 4.8\times10^{-4}, 2.4\times10^{-4},
4.3\times10^{-5}$ in {\bf (a)} and
$\chi=6.5\times10^{-2},4.5\times10^{-2},3.3\times10^{-2},
2.3\times10^{-2},1.2\times10^{-2},
4.1\times10^{-3},2.1\times10^{-3}, 1.4\times10^{-3},
7.2\times10^{-4}, 1.3\times10^{-4}, 4.5\times10^{-5}$ in {\bf (b)},
where we used the values of diffusion constant $D=0.1,0.05,0.01,0.005,0.001,0.0005$
and the Lamb-Dicke parameter $\eta=0.25,0.125,0.0625, 0.03125$.}
\label{fig6}
\end{figure*}

Here, we study how the decoherence affects the quantum-classical differences
given by the quantity ${\mathcal D}_n$, when the initial distribution
explore phase space regions of regular or mixed classical dynamics.
Our approach is in line with the one developed in Ref. \cite{Toscano2005}, where
the case of chaotic dynamics was studied.
Therefore, we also coupled the KHO to a thermal reservoir with average population
$\bar{n}$, in the Markovian and weak coupling limit, and we consider the purely diffusive regime \cite{Toscano2005}, {\it i.e.} the high temperature
regime $\bar{n}\rightarrow \infty$ together with $\Gamma \rightarrow 0$ but when
$\bar{n}\Gamma$ is constant ($\Gamma$ is the dissipation rate).
In this limit, the action of the reservoir over the Wigner function is described
by the Fokker-Planck equation
\begin{equation}
\label{Fokker_Planck}
\left. {\frac{{\partial W}}{{\partial t}}} \right|_{{\rm{reservoir}}} =
\tilde{\Gamma} \left( {\frac{{\partial ^2 W}}{{\partial q^2 }} + \frac{{\partial ^2 W}}{{\partial p^2 }}} \right)\,,
\end{equation}
with $\tilde{\Gamma}\equiv\bar{n}\Gamma\eta^2$.
For the complete evolution we have to add to right-hand side (RHS) of
Eq.(\ref{Fokker_Planck}) the unitary evolution given by the Hamiltonian in Eq.(\ref{Hamiltonian_kicked_rotor}).
The action of a pure diffusive reservoir over a classical distribution is also described by Eq.(\ref{Fokker_Planck}) if we identified $\tilde{\Gamma}$ with
the classical diffusion constant. In this case, we must add to the RHS of Eq.(\ref{Fokker_Planck}) the corresponding Liouville term
of the classical dynamics to obtain the complete evolution.

Our findings about the behavior of the quantity ${\mathcal D}_n$ in the presence
of decoherence are a consequence of the effects of the reservoir over the
one kick propagator of the Wigner function, that bring it close to the classical propagator of one kick.
Hence,  we first summarize the steps in Ref. \cite{Toscano2005} that allows to write
the one kick propagator of the Wigner function as a function of the one kick
propagator of the classical distribution under the action of a purely diffusive reservoir and in the semiclassical regime ($\eta\ll 1$).

\begin{figure*}[t]
\setlength{\unitlength}{1cm}
\includegraphics[width=14.5cm]{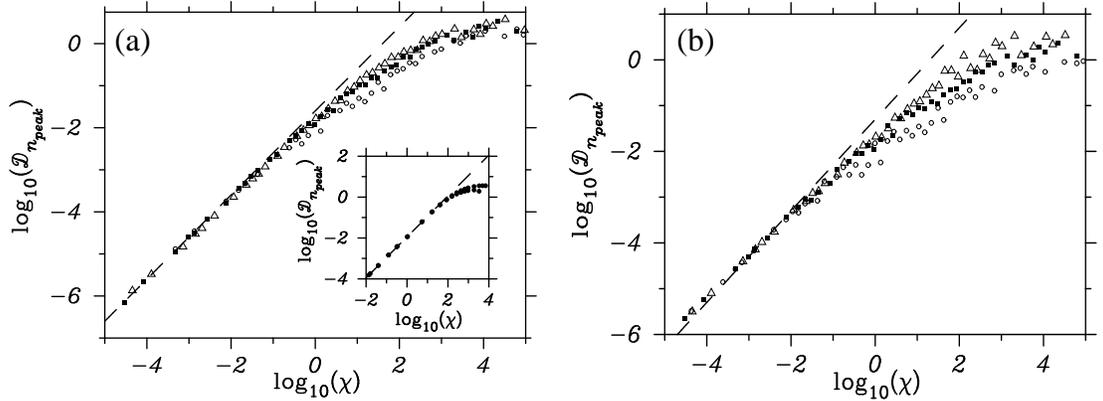}
\caption{The maximum distance between quantum and classical distributions
given by the value of the first peak of ${\mathcal D}_n$ at $n=n_{\rm peak}$
as a function of $\chi$.
The initial distribution is centered at ${\bf x}_0=(0,1.1)$ in {\bf (a)}
and ${\bf x}_0=(0.7,0)$ in {\bf (b)}.
The marks $(\circ)$ are for $K=0.5$, the $(\blacksquare)$ for $K=1.0$ and
$(\vartriangle)$ for $K=1.5$.
The region of linear behavior of ${\mathcal D}_{n_{\rm peak}}$
as a function of $\chi$ is highlighted by the dashed line with unit value of the slope.
The inset is the result borrow from \cite{Toscano2005} when the initial distribution
explore a chaotic region in phase space of the classic KHO.
We used the value of the Lamb-Dicke parameters $\eta=0.5,0.25, 0.125,0.0625,0.03125$
and the diffusion constant $D=5\times 10^{-6},
1\times 10^{-5}, 5\times 10^{-5}, 1\times 10^{-4}, 5\times 10^{-4}, 1\times 10^{-3},
5\times 10^{-3}, 1\times 10^{-2}, 5\times 10^{-2}, 10^{-1}$.}
\label{fig7}
\end{figure*}

The solution of the differential equations for the evolutions of both, the
classical and the quantum distributions, in the interval of time between two
consecutive kicks, and in the presence of the purely diffusive reservoir, can
be written as in Eq.(\ref{Wigner_evolution}) where
we have to replace the one kick propagator $L({\bf \bar{x}},{\bf x'})$ of
Eq.(\ref{quantum-propagator}) or (\ref{classical-propagator}) by the
smoothed one
\begin{equation}
\label{smoothed-propagators}
\tilde{L}({\bf x}^R,{\bf x'})\equiv\int \frac{d{\bf \bar{x}}}{2\pi D}\;
L({\bf \bar{x}},{\bf x'})
\exp\left(-\frac{({\bf \bar{x}}-{\bf x}^R)^2}{2D}\right)\;,
\end{equation}
here $D=\tilde{\Gamma}\tau$ is the diffusion constant between two consecutive
kicks, and ${\bf x}^R$ is the result of the harmonic evolution
given in Eq.(\ref{x-rotado}).
Then, the classical smoothed propagator for the classical distribution
function is
\begin{equation}
\label{smooth_classical_propagator}
\tilde{L}^{cl}({\bf x}^R,{\bf x'})=\frac{e^{-\left(x^2+y^2\right)}}{4\pi D}\;,
\end{equation}
where $y=\left( {p' - p^R  + K\sin q'} \right)/2\sqrt D$
and $x= \left( {q' - q^R } \right)/2\sqrt D$.
In the same way, we performed the integration in Eq.(\ref{smoothed-propagators})
over the the propagator in Eq.(\ref{quantum-propagator}) to obtain the smoothed quantum propagator:
\begin{widetext}
\begin{equation}
\label{smoothed-quantum-propagator}
\tilde{L}({\bf x}^R,{\bf x'})\equiv\int_{-\infty}^{+\infty}
\frac{d\mu}{2\pi \eta^2}\;e^{-\frac{D\mu^2}{\eta^4}}\;
e^{\frac{i}{\eta^2}\left[K \sin(q')\sin(\mu)-
\mu(p^R-p')\right]}\,
\delta\left(q^R-q'\right)\,.
\end{equation}
\end{widetext}
%


The presence of the Gaussian factor $\exp(-D\mu^2/\eta^4)$ in the integrand
allows to obtain an approximation in terms of the classical smoothed propagator
given in Eq.(\ref{smooth_classical_propagator}).
Indeed, when the width of this Gaussian is small, $\eta^2/\sqrt{D}\ll 1$, the $\mu$'s that effectively contribute to the integration are those close to the origin.
So, it can be used $\sin(\mu)\approx\mu-\mu^3/6$ in the phase
of the integrand.
Moreover, if $\chi=K\eta^4/D^{3/2}\ll 1$, the term with $\mu^3$ in
the phase is small, so it also can be used
$e^{i(\theta+\delta)}\approx e^{i\theta}+
i\delta\,e^{i\theta}$ and then the $\mu$-integration can be performed.
Thus, it is obtained the following approximation to the smoothed
quantum propagator:
\begin{equation}
\label{approx_smooth_quantum_propagator}
\tilde L ({\bf x}^R,{\bf x'}) \approx \tilde L^{cl}({\bf x}^R,{\bf x'})\left[1  + \chi
 \sin (q')f(y) \right]\,,
\end{equation}
where $f\left( y \right) = {1}/{4}\left( y - {2y^3 }/{3} \right)$.
Since $\left| {f\left( y \right)}exp\left(-y^2\right) \right| \le 0.081$, Eq.~(\ref{approx_smooth_quantum_propagator}) is valid under the less restrictive condition $\chi\lesssim1$.
It is important to note that the smoothed quantum propagator 
(Eq.(\ref{approx_smooth_quantum_propagator})) is obtained without any
assumption of the underlying classical dynamics of the system.
The only conditions that have to be fulfilled are:
{\bf i)} $D \gg\eta^4$ and {\bf ii)} $D \gtrsim (K\eta^4)^{2/3}$.
These two conditions  can be casted
in $D \gtrsim (K\eta^4)^{2/3}\gg\eta^4$, which is always attainable in the
semiclassical regime ($\eta \ll 1$), independently of the value of $D$.

The approximation of the quantum propagator [Eq.(\ref{approx_smooth_quantum_propagator})] allows us
to write the Wigner function immediately before the kick $n$ as,
\begin{equation}
\label{wigner-approx}
\tilde{W}_n({\bf x})=\tilde{W}^{cl}_n({\bf x})+
\chi\sum_{j=0}^{n}\,\tilde{G}_j({\bf x})+
{\cal O}(\chi^2)\;\;.
\end{equation}
where the tilde indicates the action of the purely diffusive reservoir and we
define the phase space function,
\begin{eqnarray}
\label{quantum_iteration}
\tilde{G}_{j}({\bf x})&=&
\int d{\bf x}_n \,\tilde{L}^{cl}({\bf x}^R,{\bf x}_n)
\nonumber\\
&&\vdots \nonumber \\
&\times&\int d{\bf x}_{j}\,\sin(q_j)\,f(y_j)\,
\,\tilde{L}^{cl}({\bf x}^R_j,{\bf x}_{j-1})\times\nonumber\\
&&\vdots \nonumber \\
&\times&
\int d{\bf x}_{0}\;\tilde{L}^{cl}({\bf x}^R_1,{\bf x}_{0})\;
W_0({\bf x}_0)
\,\;.
\end{eqnarray}

\noindent
Eq.(\ref{wigner-approx}) is obtained iterating
the evolution given by Eq.(\ref{Wigner_evolution}) with the 
smoothed Wigner propagator in Eq.(\ref{approx_smooth_quantum_propagator}) 
from the initial condition $W_0({\bf x}_0)$,
and keeping only terms of first order in $\chi$.
Hence, when $\chi\ll 1$ we get for the separation between the classical and
quantum distributions,
\begin{equation}
\label{Dn-prop-chi}
{\mathcal D}_n\approx\chi\,\int d{\bf x} \,
\left|\sum_{j=0}^{n}\,\tilde{G}_j({\bf x})\right|\;\;.
\end{equation}

The scaling law proportional to $\chi$ is confirmed in Fig.~\ref{fig6}
where we displayed ${\mathcal D}_n/\chi$ as a function of the number of kicks
$n$ for $K=0.5$ and $K=1.5$ and a wide range of values of $\eta$ and $D$.
We can see that for order of magnitude different values of $\chi$, all the curves
${\mathcal D}_n/\chi$.vs.$n$ fit in the same scale. So, independently of the value
of $D$, in the semiclassical limit we have $D \gg (K\eta^4)^{2/3}\gg\eta^4$, and
thus the separation between the quantum and the classical distributions goes
down with $\eta^4$, becoming arbitrarily small.
The curves in Fig.~\ref{fig6} were generated for an initial
coherent state whose Gaussian Wigner function, $W_0({\bf x}_0)$, is centered
in the phase space point ${\bf x}_0=(0,1.1)$ (Fig.~\ref{fig1}) but we get
similar results when $W_0({\bf x}_0)$ is centered at ${\bf x}_0=(0.7,0)$.
It is worth to remember that the initial distributions considered explore
regions of phase space with essentially regular classical dynamics when K=0.5,
while for K=1.5 these regions correspond to a mixed classical dynamics.
We have also obtained similar results when we considered the case with $K=1$.

The range of values of $\chi$ for which the approximation in
Eq.(\ref{Dn-prop-chi}) works is showed in Fig.~\ref{fig7} where we
plot the first peak of ${\mathcal D}_n$, that corresponds to
the maximum separation between the classical and the quantum distributions,
as a function of $\chi$.
We can see that the proportionality of ${\mathcal D}_n$ with $\chi$ is in general
valid for $\chi < 1$  with a slightly dependence with the position in phase
space of the initial distribution.
We also observe that, when the phase space regions explored by the
initial distribution are regular or mixed the proportionality
is valid up to values ${\mathcal D}_n < 1$ while
if the regions explored are essentially chaotic the proportionality with
$\chi$ is valid up to ${\mathcal D}_n\approx 1$ (see inset of Fig.~\ref{fig7}).

It is interesting to compare the behavior of the time position
of the first peak of ${\mathcal D}_n$
as a function of $\eta$ of our curves with its behavior
when $K=2$ found in \cite{Toscano2005}.
In that case the initial distribution was centered 
at the origin of phase space, that for $K=2$ it is an hyperbolic fixed 
point surrounded by a chaotic region (Fig.~\ref{fig1}(d)).
The time position of the peak occurred approximately for
$n_{\rm peak}(\eta)\approx (1/\Lambda)\ln(1/\eta)$ where
$\Lambda$ is the expansion eigenvalue of the linear map
at the origin.  However, in our case the time position
of the first peak of ${\mathcal D}_n$ is almost the same for all values of
$\eta $. 
Note that the first peak of ${\mathcal D}_n$ points 
the moment when the decoherence effects start to dominate, restraining the 
tendency to the separation of the classical and quantum distributions
produced by the presence of nonlinearities in the system.
When the underlying classical dynamics is chaotic these nonlinearities 
are reached in a very short time, so the diffusion dominates when
these nonlinearities have already been reached.
This reflects in the behavior of the position of the peak that 
it is a memory of the behavior of the separation time
Eq.(\ref{separation-time-chaotic}) without the action
of a reservoir, although lost its meaning
due to the smallness of ${\mathcal D}_n$.
On the contrary, when the underlying classical dynamics is regular
or mixed the time to reach the nonlinearities is very slow
so almost immediately the diffusion washed out any memory, in the position
of peak, of the separation time in Eq.(\ref{separation-time integrable}).

Our numerical simulations together with those in \cite{Toscano2005} show
that the scale-law  in Eq.(\ref{Dn-prop-chi}) rules the quantum to classical
transition of the KHO in the semiclassical regime
and in the presence of decoherence, independently of the underlying classical
behavior.
This is a consequence of the approximation in Eq.(\ref{wigner-approx}).
Therefore, the same scale-law proportional to $\chi$ is valid for the mean
values of any smooth observables throughout the evolution of the system.
Indeed, the mean value of an observable $\hat{O}$ in a state represented by the
Wigner function $W({\bf x})$ is given by
\begin{equation}
\langle \hat{O}\rangle = \int d{\bf x}\, O({\bf x})\, W({\bf x})\;\;\;,
\end{equation}
where $O({\bf x})$ is the Weyl symbol of $\hat{O}$ obtained through the
Weyl transform in Eq.(\ref{def-wigner-function})
replacing $\hat{\rho}/4\pi\eta^2 \rightarrow \hat{O}$ \cite{Hillery1984}.
The classical mean value is obtained in the same way
substituting $O({\bf x})$ by the classical function $O^{cl}({\bf x})$
and $W({\bf x})$ by the classical distribution $W^{cl}({\bf x})$.
So, for every smooth operator in the semiclassical regime we can take
$O({\bf x})\approx O^{cl}({\bf x})$ \cite{Alfredo1998} and in the presence
of decoherence we use the result in Eq.(\ref{wigner-approx}) to get
\begin{equation}
\langle \hat{O}\rangle_n - \langle \hat{O}\rangle_n^{cl}= \chi
\int d{\bf x}\, O^{cl}_n({\bf x})\,\sum_{j=0}^{n}\,\tilde{G}_j({\bf x})\,+\,
{\cal O}(\chi^2)\;\;.
\end{equation}


\section{Final remarks}
\label{section4}

We study the effect of decoherence in the quantum-classical transition 
of the KHO when the classical dynamics is regular or mixed (with small regions 
of chaotic behavior).
Our approach is based on the behavior of the distance ${\cal D}_n$ [Eq.(\ref{functional_distance})]
as a function of time.
We show that without the presence of a reservoir, ${\cal D}_n$ clearly indicates
that the quantum unitary evolution separates from the classical evolution
because it reaches values that do not decrease in the classical limit.
When the initial distribution explore regions of regular classical dynamics,
the separation time $t_s$ scales as a potential law 
with the effective Planck constant, that is,  $t_s \approx \tau/\hbar_{\rm eff}^{\alpha}$.
We observe that when the degree of mixing increase this 
potential-law is not valid anymore. 

When the system is in contact 
with a purely diffusive reservoir, we extend the results of Ref. 
\cite{Toscano2005}  for regular and mixed classical dynamics, 
showing that in these regimes ${\cal D}_n$ is also proportional 
to $\chi=K\hbar_{\rm eff}^2/4D^{3/2}$ when $\chi < 1$.
This implies that, in the semiclassical regime 
($\hbar_{\rm eff} \ll 1$), and independently of the values of 
the diffusion constant $D$ and the classical dynamics explored by the 
initial distribution, the quantum-classical differences
${\cal D}_n$ goes to zero as $\hbar_{\rm eff}^2$.
The conditions for the proportionality between ${\cal D}_n$ and $\chi$
can be condensed in the inequality 
$D \gg (K\hbar_{\rm eff}^2)^{2/3}\gg \hbar_{\rm eff}^2$
that is the relation of the strength of noise and the semiclassical 
parameter for the KHO in order to obtain
correctly the classical limit for all the regimes of its underlying classical
dynamics. 
Finally, we can say that because of the smallness of  ${\cal D}_n$, 
in the semiclassical regime, the concept of separation time 
is not meaningful anymore.

Although in the framework of an specific model, our results and
those in Ref. \cite{Toscano2005}, indicates that
the conjecture presented in Ref. \cite{Pat}, that in the presence of noise 
a single parameter $\chi$ controls the quantum-classical transition 
of classically chaotic systems, may be extended
in general for all type of classical dynamics.
However, it seems that in $\chi$ should enters
only a parameter that controls the degree of mixing of the considered system  
and not directly the Lyapunov coefficient that
characterize the exponential divergence of trajectories in fully chaotic 
classical systems.


\section{Acknowledgments}
We acknowledge Luiz Davidovich for fruitful 
comments. 
DAW gratefully acknowledges support from CONICET(Argentina) and PROSUL(Brazil). 
FT acknowledges the hospitality at the
"Departamento de F\'{\i}sica ''J. J. Giambiagi''", FCEN, UBA,
where this work started.







\end{document}